\documentclass[aps,prb,reprint,longbibliography,superscriptaddress]{revtex4-1}

\usepackage{hyperref}
\usepackage{wrapfig}
\usepackage{graphicx}
\usepackage{epstopdf}
\usepackage{amssymb}
\usepackage{amsmath}
\usepackage{bbm}
\usepackage{color}
\usepackage{xfrac}
\usepackage{datetime}
\usepackage{braket}

\newlength{\tlength}

  \renewcommand{\Im}{\mathop{\rm Im}}
  \newcommand{\Tr}{\mathop{\rm Tr}\nolimits}
  \newcommand{\e}{\mathrm{e}}
  \let\ifr\i
  \renewcommand{\i}{{\rm i}}

  \renewcommand{\d}{{\rm d}}

\begin{document}

\title{Hole-capture competition between a single quantum dot and an ionized acceptor}
\date{\today}

\author{J. Wiegand}
\thanks{These authors contributed equally to this work.}
\affiliation{Institute for Solid State Physics, Leibniz Universit\"at Hannover, Appelstr. 2, 30167 Hannover, Germany}
\affiliation{Laboratory of Nano and Quantum Engineering (LNQE), Leibniz Universit\"at Hannover, Schneiderberg 39, 30167 Hannover, Germany}
\author{D. S. Smirnov}
\thanks{These authors contributed equally to this work.}
\email{smirnov@mail.ioffe.ru}
\affiliation{Ioffe Institute, Polytechnicheskaya 26, 194021 St. Petersburg, Russia}
\author{J. Osberghaus}
\author{L. Abaspour}
\author{J. H\"ubner}
\author{M. Oestreich}
\affiliation{Institute for Solid State Physics, Leibniz Universit\"at Hannover, Appelstr. 2, 30167 Hannover, Germany}
\affiliation{Laboratory of Nano and Quantum Engineering (LNQE), Leibniz Universit\"at Hannover, Schneiderberg 39, 30167 Hannover, Germany}

\begin{abstract}
We study the competition of hole capture between an In(Ga)As quantum dot and a directly adjacent ionized impurity in view of spin-photon interfaces. The Kerr rotation noise spectroscopy at 4.2~K shows that the hole-capture probability of the In(Ga)As quantum dot is about one order of magnitude higher compared to the hole-capture probability of the ionized impurity and suggests that a simultaneous occupation of quantum dot and impurity by a hole is efficiently suppressed due to Coulomb interaction. A theoretical model of interconnected spin and charge noise allows the quantitative specification of all relevant time scales.
\end{abstract}

\maketitle

\section{Introduction}
 
The spin of a single hole localized in a semiconductor quantum dot (QD) is envisioned as qubit for prospective spin-photon interfaces.\cite{Arnold.2015} The necessary hole is conveniently provided in In(Ga)As QDs by the typical p-type background doping due to carbon impurities. This circumstance avoids the need of complex electrical gate structures and simplifies experiments on spin-photon interfaces enormously. Hole spins in In(Ga)As QD based spin devices outperform electron spins since the p-type heavy-hole Bloch wavefunction diminishes spin relaxation via hyperfine interaction in comparison to the s-type Bloch wavefunction of electrons in the conduction band.\cite{Heiss.2007,Chekhovich.2013}

Single QD spin-photon interfaces\cite{Gao.2015,Press.2008,Imamoglu.1999,Mi.2018} will typically be driven quasi-resonantly at the trion resonance which effectively prevents the inauspicious creation of free carriers. However, the excited trion state has a very small but nevertheless finite probability to undergo an Auger process\cite{Kurzmann.2016} which excites the information-carrying hole out of the QD into the valence band continuum. This process has been observed by Kerr rotation noise spectroscopy in an In(Ga)As QD recently,\cite{Wiegand.2018} and interrupts the functional capability of the spin-photon device until a hole is recaptured by the QD. In fact, the Auger process can start a competition of hole capture between the single QD and the adjacent, ionized acceptor which provided the hole initially. The QD has the advantage of its larger size and a very deep trapping potential while the ionized acceptor is negatively charged and attracts the hole efficiently by the Coulomb force. Hence, the ability to measure the different capture and acceptor-to-QD tunneling rates is important for the design of QD spin-photon interfaces. 

In the following we present Kerr rotation noise measurements on a single In(Ga)As QD that reveal the presence of a single acceptor in the direct vicinity of the QD. The acceptor provides a single hole to the QD by hole tunneling. We will show that in this case the Kerr rotation noise spectrum consists of three different contributions. The first contribution is related to the spin dynamics in the ground and optically excited trion states, which is commonly known as spin noise spectroscopy (SNS).\cite{Zapasskii:13,Oestreich-review,SinitsynReview} A second contribution results from the Auger initiated excitation of the hole into the valence band and a subsequent capture of a hole directly by the QD. The third contribution is also related to the Auger process but followed by the capture of a hole at the negatively charged, ionized acceptor, and a subsequent slow tunneling process of the hole from the impurity into the QD.

The paper is structured as follows: We start with the presentation of the experimental results in section~\ref{s:exp}. A phenomenological model and the corresponding theory are described in sections~\ref{s:model} and~\ref{s:theory}. In section~\ref{s:dis} the theoretical model is applied to analyze the experimental results followed by a conclusion in section~\ref{s:con}.

\section{\label{s:exp}Experiment}

\begin{figure*}
\centering
\includegraphics[width=\linewidth]{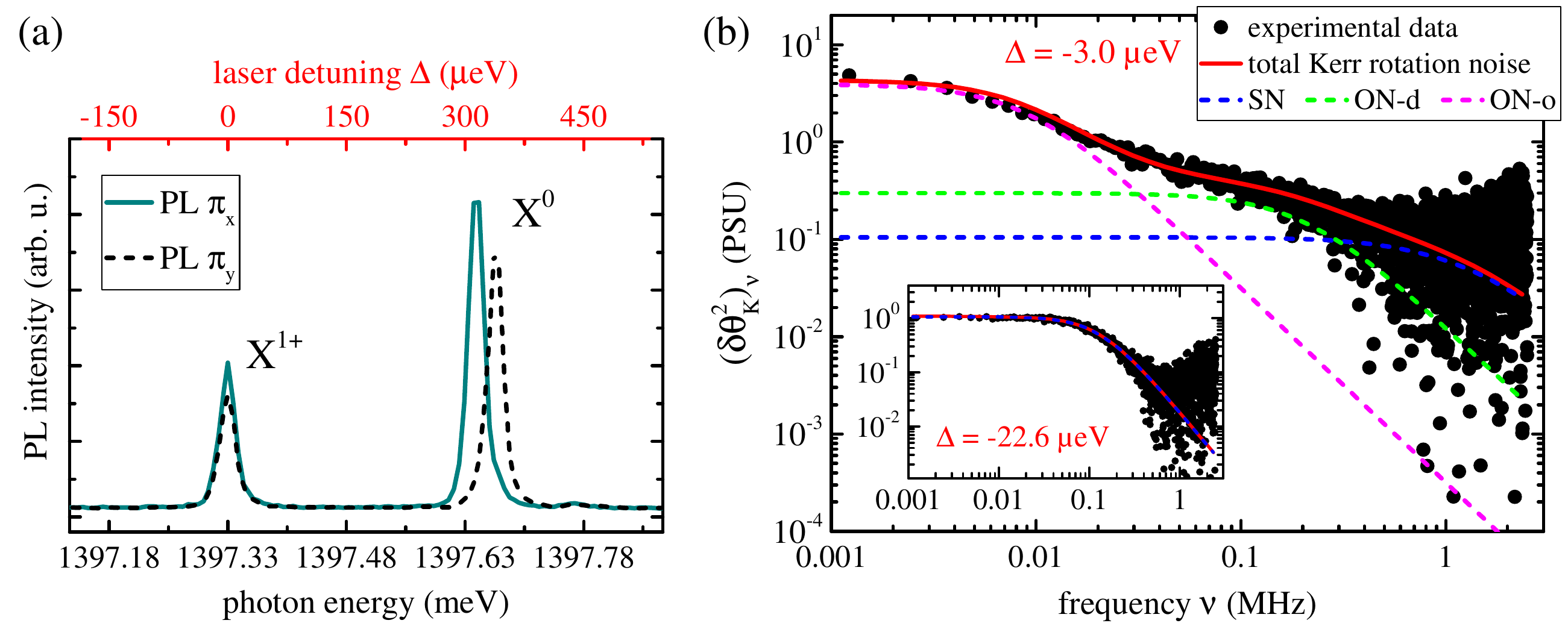} 
\caption{\label{fig:spectra}(a) Polarization-resolved PL spectrum of the QD revealing trion ($\rm X^{1+}$) and exciton ($\rm X^0$) transition. The observed energetic splitting of the PL of $X^0$ for the two perpendicular linear polarization directions is typical for an uncharged In(Ga)As QD state. The measured PL linewidth is mainly limited by the spectrometer resolution of about 20~$\mu$eV. Kerr rotation noise measurements are performed with respect to the $\rm X^{1+}$ transition with a probe laser detuning $\Delta$. (b) Kerr rotation noise spectrum in units of the photon shot noise (PSU) measured at a small detuning of $\Delta=-3\ \mu$eV. The noise spectrum deviates significantly from the pure SN spectrum (dashed blue line, SN). At large detunings, as shown in the inset ($\Delta=-22.6\ \mu$eV), the additional noise contributions vanish and the spectrum is fully described by SN. The red (blue) lines correspond to the total Kerr rotation noise (spin noise) spectrum calculated by the theoretical model including occupation noise of QD \mbox{(ON-d)} and outer state \mbox{(ON-o)}, i.e., according to Eq.~\eqref{eq:total_spectrum} with parameters summarized in Tab.~\ref{tab:params}.}
\end{figure*}

\begin{figure*}[t]
\centering
\includegraphics[width=\linewidth]{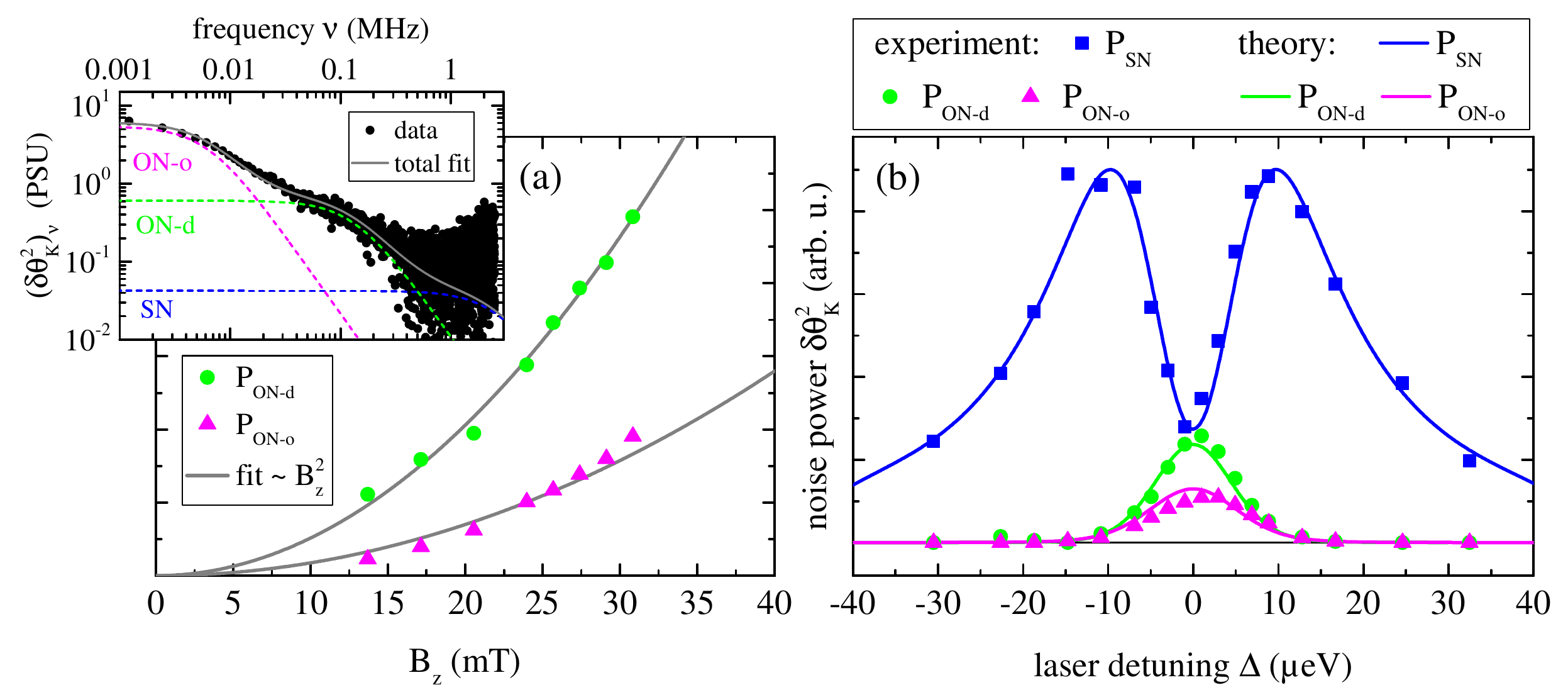}
\caption{\label{fig:power}(a) The inset shows the Kerr rotation noise spectrum measured at a laser detuning $\Delta = 1\ \mu$eV. The dashed lines illustrate the phenomenological fit of three Lorentzian contributions to the spectrum. The main panel shows the noise power related to the area of the Lorentzian contributions \mbox{ON-d} and \mbox{ON-o} as a function of $B_z$ for $\Delta = 1\ \mu$eV. 
(b) The noise power of SN and ON contributions as a function of laser detuning. The symbols depict the respective noise powers extracted from the phenomenological fit to the measured spectra. The solid lines correspond to the noise power contributions calculated by Eq.~\eqref{eq:power_spectra} of the theoretical model using parameters in Tab.~\ref{tab:params}.}
\end{figure*}

The sample comprises a single layer of self-assembled In(Ga)As QDs grown by molecular beam epitaxy on a (001)-oriented GaAs substrate. The QD layer has a gradient in QD density from zero to about 100~dots/$\mu\rm{m}^2$. An unintentional p-type background doping of $\approx 10^{14}$cm$^{-3}$ ensures that a fraction of the QDs is occupied by a single hole. The QDs are embedded in an asymmetric GaAs $\lambda$ microcavity with 13 (top) and 30 (bottom) AlAs/GaAs Bragg mirror pairs giving a Q-factor of about 350. The measurement setup is a low-temperature confocal microscope with two detection arms, one for photoluminescence (PL) analysis and one for Kerr rotation noise spectroscopy (see Refs.~\onlinecite{Dahbashi.2012,Wiegand.2018} for details). The QD sample is cooled down to 4.2~K and, for PL measurements, excited above the QD barrier with a cw diode laser at a photon energy of 1.59~eV. The solid and dashed lines in Fig.~\ref{fig:spectra}(a) show the polarization-resolved PL spectra of a single QD in a sample region with  low QD density. The optical transition at higher energy exhibits a small fine structure splitting between the perpendicular polarized exciton eigenstates, $\pi_x$ and $\pi_y$, induced by the anisotropic exchange interaction.\cite{Bayer.2002} This interaction is absent and the splitting between the PL components vanishes if the QD is occupied by an additional (resident) hole provided by the p-type doping. Hence, the high-energy transition in the PL spectrum corresponds to the neutral exciton ($X^0$) and the low-energy transition is attributed to the positively charged trion ($X^{1+}$), which consists of the resident hole plus an optically excited electron-hole pair. The assignment is confirmed by the following spin noise measurements. We selected this specific QD since Kerr rotation occupation noise measurements show for this QD that one of the stochastically distributed acceptors of the background doping is so close to the QD that the hole tunneling time is in the easily measurable sub-millisecond regime. All results below are measured on this specific QD. 

Kerr rotation noise of a single QD has been measured only twice so far.\footnote{Both measurements were done on the same sample as used in this work} The very first measurement has been carried out on a strongly inhomogeneously broadened QD\cite{Dahbashi.2014} and the second measurement on a homogeneously broadened QD\cite{Wiegand.2018} which additionally revealed hole occupation noise (ON) due to the Auger process. However, hole dynamics directly related to acceptors transitions have not been observed. The current measurement focuses on a mainly homogeneously broadened QD where an acceptor is in close vicinity (on the order of a few Bohr radii of the acceptor hole, $a_B^{\rm{acc}} \approx 2$~nm~\cite{baldereschi73} 
), such that QD and acceptor compete for the same hole and Coulomb interaction effectively suppresses simultaneous occupation of acceptor and QD. 

All Kerr rotation noise measurements presented in the following are carried out in a longitudinal magnetic field $B_z=31$~mT and with a laser power of 1.5~$\mu$W focused to about 1~$\mu$m unless otherwise noted. 
The probe laser is a cw Ti:sapphire ring laser which is stabilized to a Fizeau wavelength meter. The Kerr rotation noise due to the single QD is imprinted onto the linear laser polarization and Fourier transformed in real time to obtain a Kerr rotation noise spectrum.

The inset of Fig.~\ref{fig:spectra}(b) shows the Kerr rotation noise spectrum of the QD in units of the photon shot noise measured with a large laser detuning from the $X^{1+}$ transition of $\Delta=-22.6\ \mu$eV. This well detuned Kerr rotation noise spectrum is dominated by spin noise (SN) which has a pure single Lorentzian line shape. The half width at half maximum (HWHM) of the SN spectrum is proportional to a spin relaxation rate, which is the hole-spin relaxation rate if optical excitation by the probe laser is negligible. However, strong resonant excitation leads to the formation of coherent superpositions of hole and trion states, called ``dressed states'', and the corresponding SN spectrum is determined by the dressed-state spin relaxation which strongly depends on detuning and laser intensity.\cite{Wiegand.2018} The noise spectrum at $\Delta=-22.6\ \mu$eV yields a spin relaxation time of about $1\;\mu$s. In fact, even at this large detuning the spin relaxation of the dressed state is not dominated by the hole spin but mainly by the electron-in-trion spin flips (see Ref.~\onlinecite{Wiegand.2018} and sec.~\ref{s:theory}). The intrinsic QD hole-spin relaxation time $T_1^h$ is only observed at even larger detunings and/or lower laser intensities and is so long that it doesn't play any role in this experiment.\cite{Dahbashi.2014} This is in contrast to a free hole in the valence band continuum where the spin relaxation is on the order of the momentum relaxation time, i.e., all holes in the continuum loose their spin quasi instantaneously. 

Figure~\ref{fig:spectra}(b) shows in contrast the Kerr rotation noise spectrum for the same laser intensity but a small detuning of $\Delta=-3\ \mu$eV. This spectrum consists of three clearly distinguishable Lorentzian contributions. The broadest Lorentzian (blue, dashed curve) results from the SN of the strongly driven QD transition with a spin relaxation time of about 140~ns. This spin relaxation time $\tau_s$ is about one order of magnitude faster than $\tau_s$ at large detunings due to the increased probability of trion population, i.e., stronger influence of electron spin relaxation in conjunction with the fast Rabi oscillations between ground and trion states.
We can unambiguously assign this Lorentzian contribution to SN since its noise power as a function of detuning, $P_{\rm SN}(\Delta)$, exhibits the typical SN lineshape of a nearly homogenously broadened QD transition with a clear dip at zero detuning (see blue curve and squares in Fig.~\ref{fig:power}(b)).\cite{Zapasskii.2013} 

In the following we will focus on the other two Kerr rotation noise contributions which become dominant in the low-frequency regime at small laser detunings as shown in the inset of Fig.~\ref{fig:power}(a) for the detuning of $\Delta=1\;\mu$eV. The additional noise contributions are labeled \mbox{ON-d} and \mbox{ON-o}, and analysis of their respective noise powers, $P_{\text{ON-d}}$ and $P_{\text{ON-o}}$, as a function of laser detuning and external magnetic field enables to identify their origin. The noise powers are extracted from the area of the respective Lorentzian fits. Figure~\ref{fig:power}(a) shows that the noise power of both contributions increases approximately by the square of the external magnetic field $B_z$, where the $z$-direction is the QD growth and the noise-detection direction. Additionally, Fig.~\ref{fig:power}(b) shows that the noise power of both contributions exhibits a maximum at the trion resonance. Both observations together prove that \mbox{ON-d} and \mbox{ON-o} result from occupation noise.\cite{Wiegand.2018}\textsuperscript{,}\footnote{The increase of the noise power with the square of $B_z$ is in our case a good approximation whereas higher orders play an important role e.g. for higher $B_z$.} 

\section{\label{s:model}Model}

\begin{figure*}
\centering
\includegraphics[width=\linewidth]{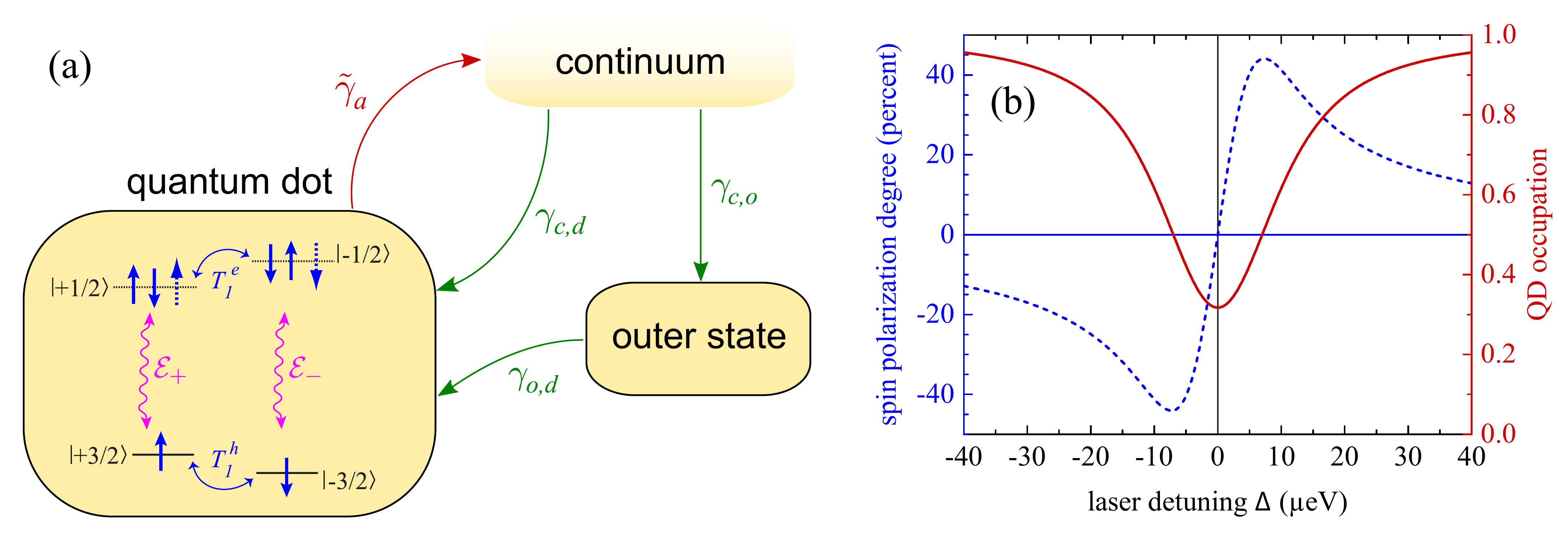}
\caption{\label{fig:model}(a) Sketch of the QD states and of the hole states outside of the QD. The arrows depict the relevant transitions between the states. (b) Degree of spin polarization due to unequal optical pumping of the two Zeeman levels according to Eq.~\eqref{eq:P} (blue dashed line) and the average QD occupation according to Eq.~\eqref{eq:nav} (red solid line) as functions of detuning. Parameters of the calculation are given in Tab.~\ref{tab:params}.}
\end{figure*}

The microscopic origin of QD occupation noise, or noise of the QD charge status, is related to trion Auger recombination which expels the resident hole out of the QD. After some time a free hole is captured by the empty QD whereat the exact capture process and time are determined by the solid-state environment. The external magnetic field leads to a dc Kerr rotation when the QD is charged (and the probe light is quasi-resonant with the trion transition). Emptying of the QD results in disappearance of this rotation angle, and thereby to Kerr rotation noise. In order to quantitatively describe the Kerr rotation noise frequency and power spectra we develop a model of coupled charge and spin dynamics in the QD.
Figure~\ref{fig:model}(a) depicts schematically the underlying scenario. The positively charged QD in the presence of probe light can be described by a four-level system.\cite{glazov.2016,Smirnov.2017} The ground states are $\ket{\pm3/2}$ with the corresponding projection of the heavy-hole spin on the growth axis. Absorption of a photon by the QD leads to the formation of singlet trion states $\ket{\pm1/2}$, which are characterized by the spin of the electron in the trion. According to optical selection rules, $\sigma^\mp$-polarized light couples the states $\ket{\pm3/2}$ and $\ket{\pm1/2}$, respectively.\cite{Ivchenko.op.2005,Dyakonov.op.2017} The spin dynamics in the QD are characterized by the hole-spin relaxation time $T_1^h$ in the ground state and by the electron-spin relaxation time $T_1^e$ in the excited state whereat $T_1^h \gg T_1^e$ and can be neglected.

\begin{table}
\centering
\begin{tabular}{|l|l|l|}
\hline
physical quantity & symbol & value \\  
\hline\hline
``dressed'' homogeneous trion linewidth & $\gamma_1$ & 4.3~$\mu$eV \\
\hline
inhomogeneous broadening & $\hbar\Delta\omega_0$ & 4.5~$\mu$eV \\
\hline
effective Zeeman splitting at $B_z=31$~mT& $\hbar\Omega_z$ & 5.2~$\mu$eV \\
\hline
electron spin relaxation time & $T^e_1$ & 24~ns \\
\hline
Auger recombination rate & $\gamma_a$ & 2~$\mu$s$^{-1}$ \\
\hline
QD reoccupation rate from continuum&  $\gamma_{c,d}$ & 0.7~$\mu$s$^{-1}$ \\
\hline
outer-state reoccupation rate from cont.& $\gamma_{c,o}$ & 0.06~$\mu$s$^{-1}$ \\
\hline
QD reoccupation rate from outer state &  $\gamma_{o,d}$ & 0.03~$\mu$s$^{-1}$ \\
\hline
\end{tabular}
\caption{\label{tab:params}Parameters of spin and charge dynamics determined by comparison of the theoretical calculations and experimental results.}
\end{table} 

The trion recombination initiates an Auger process which excites the resident hole into the continuum  with a rate $\tilde{\gamma_a}$ which is a function of the trion population. After Auger recombination a hole from the continuum--with arbitrary spin orientation--is returned to the QD. The observation of two Lorentzian contributions to the occupation noise suggests that reoccupation of the QD takes place via two different channels. The first recharging channel is linked to the broader Lorentzian \mbox{ON-d} which implies a fast recharging rate. This process is modeled by direct recapturing of a hole from the continuum by the QD with the rate $\gamma_{c,d}$, where the indices $c$ and $d$ stand for continuum and (quantum) dot, respectively. The second recharging channel is characterized by a significantly slower rate linked to the narrow Lorentzian \mbox{ON-o}, which we model with the presence of a localized outer state in the close vicinity of the QD. This outer state is most likely related to an ionized acceptor, that provided a hole for the QD initially. In the case of an empty QD--after Auger recombination--a hole can be captured first from the continuum into the outer state with the rate $\gamma_{c,o}$ followed by a slow tunnel process from the outer state into the QD with the rate $\gamma_{o,d}$ ($o$ denotes the outer state). The state of the system where both, QD and outer state, are charged is forbidden in our model due to their Coulomb repulsion. If this assumption is released, the modeled Kerr rotation noise contains additional contributions from fluctuating correlations between the outer state and the QD which is not observed in the measured spectra.
In principle, the observed occupation noise could also result from a different scenario, e.g., two acceptors as outer states with two different capture and tunneling rates. However, such a scenario is improbable at such a low background doping concentration since both acceptors have to be very close to the QD to exhibit reasonable tunneling times. 

\section{\label{s:theory}Theory}

In the following we will outline the theory describing the spin and charge noise in our system based on the model described above.
According to the optical selection rules the pseudospin $S_z=\pm1/2$ is conserved during trion excitation and recombination, i.e., at timescales on the order of~$1$~ns. The fast optical processes can be separated from the slow dynamics corresponding to the electron and hole spin relaxation, trion Auger recombination, and QD recharging. Thereby, the four-level system is always in a quasi-steady state which is characterized by QD occupancy $n$ and pseudospin $S_z$. We will first calculate the quasi-steady state and afterwards derive the spin and charge noise spectra of the QD.

\subsection*{Quasi-steady state}

The Hamiltonian of the QD interacting with the probe light has the form:\cite{Smirnov.2015,Smirnov.2017b}
\begin{align} 
  \label{eq:H}
  \mathcal H=&\hbar\sum_\pm\biggl[
\omega_0 a_{\pm1/2}^\dag a_{\pm1/2}
\pm\frac{1}{2}\Omega^ha_{\pm3/2}^\dag a_{\pm3/2} \\
&\pm\frac{1}{2}\Omega^ea_{\pm1/2}^\dag a_{\pm1/2}
+ \left(\mathcal E_\pm\e^{-\i\omega t}a_{\pm 1/2}^\dag a_{\pm 3/2} +{\rm h.c.}\right)
 \biggr]\:.\nonumber 
\end{align}
Here, $a_s$ ($a_s^\dag$) with $s=\pm3/2,\pm1/2$ are the annihilation (creation) operators for the states $\ket{s}$, $\omega_0$ is the trion resonance frequency, $\hbar\Omega_z^{e(h)}=\mu_B g_z^{e(h)} B_z$ is the electron (hole) Zeeman splitting in the external magnetic field with $g_z^{e(h)}$ being the longitudinal $g$-factor, and $\mathcal E_\pm$ are the trion optical transition matrix elements in $\sigma^\mp$ polarizations. For linearly polarized probe light $\mathcal E_\pm$ is equal to $\mp\mathcal E/\sqrt{2}$, where $\mathcal E$ is proportional to the amplitude of the incident field and the optical transition dipole moment in the QD. In the following we assume $\mathcal E$ to be real. The Hamiltonian in Eq.~\eqref{eq:H} treats the electromagnetic field classically, which is valid since QD and microcavity are in the  weak coupling regime. 

The eigenstates of the Hamiltonian [Eq.~\eqref{eq:H}] are functions of the laser detuning $\Delta=\omega_0-\omega$. In fact, the detunings for $\sigma^-$ and $\sigma^+$ polarized light, $\Delta_+$ and $\Delta_-$, respectively, are different due to different $g$-factors of electrons and holes by $\Omega_z=\Omega_z^e-\Omega_z^h$ and are given by $\Delta_\pm=\Delta\pm\Omega_z/2$. For small detunings ($\Delta<\mathcal E$) strong coherent excitation of the QD leads to the formation of dressed states.\cite{CohenTannoudji.1977,Compagno.2005} The lower and upper dressed states are the eigenstates of the Hamiltonian in Eq.~\eqref{eq:H} and are given by
\begin{eqnarray}
\label{eq:dressed}
  \Psi_l^\pm & = & \cos\theta_\pm\ket{\pm3/2}+\sin\theta_\pm\ket{\pm1/2}\e^{-\i\omega t},
  \nonumber \\
  \Psi_u^\pm & = & -\sin\theta_\pm\ket{\pm3/2}+\cos\theta_\pm\ket{\pm1/2}\e^{-\i\omega t},
\end{eqnarray}
respectively, where $\theta_\pm\in\left[-\pi/4;\pi/4\right]$ is determined by
\begin{equation}
  \tan\left(2\theta\right)=-2\mathcal E_\pm/\Delta_\pm. 
\end{equation}
In case of large detunings, $\Delta\gg\mathcal E$, $\theta$ approaches $0$ and the states $\Psi^\pm_{l,u}$ become simply the hole and the trion states. The Hamiltonian [Eq.~\eqref{eq:H}] is diagonal in the basis of the dressed states [Eq.~\eqref{eq:dressed}] with the eigenvalues:
\begin{eqnarray}
  E_l^\pm & = & \pm\frac{\Omega_z^h}{2}+\frac{\Delta_\pm-{\rm sign}(\Delta_\pm)\sqrt{\Delta_\pm^2+4\mathcal E_\pm^2}}{2}, \nonumber
  \\
  E_u^\pm & = & \pm\frac{\Omega_z^h}{2}+\frac{\Delta_\pm+\sqrt{\Delta_\pm^2+4\mathcal E_\pm^2}}{2},
\end{eqnarray}
in agreement with the Floquet theorem.

The incoherent processes, i.e. dephasing and recombination, can be taken into account in the density matrix formalism. The master equation for the density matrix $\rho$ reads
\begin{equation}
  \dot{\rho}=-\i\left[\mathcal H,\rho\right]+\mathcal L\left\lbrace\rho\right\rbrace,
  \label{eq:rho}
\end{equation}
where $\mathcal L\left\lbrace\rho\right\rbrace$ is the Lindblad superoperator. The superoperator has the form
\begin{eqnarray}
  \mathcal L\left\lbrace\rho\right\rbrace=\sum_{\pm}\biggl[
\frac{\gamma_0}{2}\left(2 d_\pm\rho d_\pm^\dag-d_\pm^\dag d_\pm\rho-\rho d_\pm^\dag d_\pm\right) \nonumber \\ 
 +\frac{\gamma}{4}\left(2 d_\pm'\rho d'_\pm-d_\pm'^2\rho-\rho d_\pm'^2\right)
\biggr],
\end{eqnarray}
where $\gamma_0$ and $\gamma$ are the trion recombination and dephasing rates, $d_\pm=a_{\pm 3/2}^\dag a_{\pm 1/2}$, and $d_\pm'=a_{\pm 1/2}^\dag a_{\pm 1/2}-a_{\pm3/2}^\dag a_{\pm3/2}$.

The QD occupancy is given by $n=\Tr\left(\rho\right)$ and the pseudospin is given by $S_z=\Tr\left[\left(d_+'^2-d_-'^2\right)\rho\right]/2$. These parameters are conserved by the master equation~\eqref{eq:rho} and parametrize the quasi-steady state. The corresponding density matrix $\tilde\rho(n,S_z)$ can be readily found from Eq.~\eqref{eq:rho} and allows us to calculate in particular the occupancies of the hole and trion states $n_s$ with $s=\pm3/2,\pm1/2$. The degree of the optical QD excitation can be characterized by the probability of trion excitation:
\begin{equation}
  \label{eq:n_tr}
  \frac{n_{\pm1/2}}{n_{\pm3/2}+n_{\pm1/2}}=\frac{\mathcal E^2\left(\gamma+\gamma_0/2\right)}{\gamma_0(\gamma_1^2+\Delta_\pm^2)},
\end{equation}
where the renormalized (``dressed'') optical transition linewidth is given by
\begin{equation}
  \gamma_1=\left(\gamma+\frac{\gamma_0}{2}\right)\sqrt{1+\frac{I}{I_0}},
\end{equation}
with $I$ being the light intensity and $I_0$ being the saturation intensity, which is determined by $I/I_0=4\mathcal E^2/\left[\gamma_0(2\gamma+\gamma_0)\right]$. Note, that the trion excitation probability depends on the detuning and thus on the mutual orientation of magnetic field and QD pseudospin.

The concept of dressed states finally gives two results: The first is the renormalization of the trion homogeneous linewidth. This effect was studied in detail in Ref.~\onlinecite{Wiegand.2018} and can be seen here in Fig.~\ref{fig:power}(b). Indeed the widths of the Kerr rotation noise power spectra is in this experiment on the order of a few $\mu$eV, while the homogeneous trion linewidth in equilibrium is nearly an order of magnitude smaller.\cite{Kuhlmann.2013} The second result, obtained using the concept of dressed states, is the expression for the trion excitation probability in Eq.~\eqref{eq:n_tr}. Ultimately this expression determines the effective Auger rate $\tilde{\gamma_a}$ and thus the occupancy noise power spectrum and the widths of the corresponding Lorentzians in the Kerr rotation noise frequency spectra.

Optical spin signals are determined by the average of $d_y=-\i\left(d_++d_-\right)/\sqrt{2}$. The Faraday rotation of the probe beam is given by the component of $d_y$ which is orthogonal in phase with the electric field inside the microcavity, i.e. $\Im\Tr\left[\e^{\i\omega t}d_y\tilde\rho(n,S_z)\right]$. For simplicity we assume that the same quantity determines the Kerr rotation.\cite{Yugova.2009} As a result we find that~\cite{Wiegand.2018}
\begin{align}
  \label{eq:Kerr}
  \theta_K\propto &\frac{\mathcal E}{(\gamma_1^2+\Delta_+^2)(\gamma_1^2+\Delta_-^2)}\times \\
  &\left[4\Delta(\gamma_1^2-\frac{\Omega_z^2}{4}+\Delta^2)S_z+\Omega_z(\gamma_1^2+\frac{\Omega_z^2}{4}-\Delta^2)n\right]\:. \nonumber 
\end{align}
This expression shows that both, fluctuations of the pseudospin $S_z$ and the QD occupancy $n$, lead to fluctuations of the Kerr rotation.

\subsection*{Spin and occupancy noise spectra}

In the following we use kinetic theory to describe spin and charge fluctuations based on the phenomenological model introduced in Fig.~\ref{fig:model}(a). It is convenient to introduce the occupancies of the pseudospin-up and pseudospin-down states as $n_\pm=n_{\pm3/2}+n_{\pm1/2}=n/2\pm S_z$. As a result the kinetic equations take the form
  \begin{align}
    \dot{n}_{\pm}=& \frac{n_{\mp3/2}-n_{\pm3/2}}{2T_1^h}
    +\frac{n_{\mp1/2}-n_{\pm1/2}}{2T_1^e}\nonumber \\
    & -\gamma_an_{\pm1/2}+\frac{\gamma_{c,d}}{2}(1-n-n_o)+\frac{\gamma_{o,d}}{2}n_o,
  \end{align}
where $n_s$ ($s=\pm3/2,\pm1/2$) are determined by Eq.~\eqref{eq:n_tr} and $n_o$ is the occupancy of the outer state. Equivalently these equations can be rewritten as
\begin{align}
  \label{eq:kinetic}
  \dot{S_z}&=-\frac{S_z}{\tilde T_1}+\lambda n\:, \nonumber \\
  \dot{n}&=-\tilde\gamma_a n-\lambda' S_z+\gamma_{c,d}(1-n-n_0)+\gamma_{o,d} n_o\:,
\end{align}
where we have introduced the average spin relaxation time $\tilde T_1$, the effective Auger rate $\tilde\gamma_a$, and coupling parameters between QD pseudospin and occupancy, $\lambda$ and $\lambda'$, as 
\begin{eqnarray}
  \frac{1}{\tilde T_1}& = & \frac{1}{T_1^h}+\left(\frac{1}{T_1^e}-\frac{1}{T_1^h}+\gamma_a\right)\frac{\gamma_1^2+\Omega_z^2/4+\Delta^2}{2\Gamma^2},
  \nonumber \\
  \tilde\gamma_a & = & \gamma_a\frac{\gamma_1^2+\Omega_z^2/4+\Delta^2}{2\Gamma^2}, \\
  \lambda& = & \left(\frac{1}{T_1^e}-\frac{1}{T_1^h}+\gamma_a\right)\frac{\Delta\Omega_z}{4\Gamma^2},
  \quad
  \lambda'=-\gamma_a\frac{\Delta\Omega_z}{\Gamma^2}, \nonumber
\end{eqnarray}
with $\Gamma = \sqrt{(\gamma_1^2+\Delta_+^2)(\gamma_1^2+\Delta_-^2)}/\gamma_1$.
Here, we have assumed the realistic limit $I\gg I_0$.\cite{Wiegand.2018} One can see from the kinetic Eqs.~\eqref{eq:kinetic} that resonant excitation of the QD in an external magnetic field leads to a spin polarization, and that the rate of emptying the QD depends on its pseudospin.

The pseudospin relaxation is dominated by electron-spin flips in our experiment. Therefore the average spin relaxation rate and pseudospin pumping rates can be simply expressed as
\begin{equation}
  \frac{1}{\tilde T_1}=\frac{\gamma_1^2+\Omega_z^2/4+\Delta^2}{2T_1^e\Gamma^2},
  \qquad
  \lambda=\frac{\Delta\Omega_z}{4T_1^e\Gamma^2}\:.
\end{equation}
As a consequence the steady-state spin polarization degree is
\begin{equation}
  \label{eq:P}
  P=\frac{2\bar S_z}{\bar n}=\frac{\Delta\Omega_z}{\gamma_1^2+\Omega_z^2/4+\Delta^2}\:,
\end{equation}
where $\bar S_z$ and $\bar n$ are the average spin and occupation of the QD. We recall that the spin polarization arises under illumination of the QD by linearly polarized probe light, provided the Zeeman energy is comparable with the QD linewidth~\cite{Brossel.1952}. 

Fluctuations of pseudospin and occupation of the QD lead to fluctuations of the Kerr rotation of the probe light in agreement with Eq.~\eqref{eq:Kerr}. The Kerr rotation noise power is given by the average $\braket{\delta\theta_K^2}$ with the correlation functions of $\delta S_z$ and $\delta n$
\begin{align}
  \label{eq:avers}
  &\braket{\delta S_z^2}=\frac{\bar n}{4}-\bar S_z^2\:, \nonumber \\
  &\braket{\delta n^2}=\bar n\left(1-\bar n\right), \\
  &\braket{\delta S_z\delta n}=\braket{\delta n\delta S_z}=\bar S_z\left(1-\bar n\right)\:. \nonumber
\end{align}
Note, that the correlators in the two last lines contribute to the Kerr rotation noise spectrum only in a longitudinal magnetic field, when the two contributions to the Kerr rotation $\delta\theta_K\propto\delta S_z,B_z\delta n$ are symmetry allowed.

The external magnetic field strongly couples QD occupancy and spin polarization. The experimental results suggest that their coupled dynamics consist of three well separated timescales: the fastest process is related to the electron-spin relaxation time $T_1^e$. Auger recombination and direct QD recharging from the continuum define the intermediate timescale, while hole capture to the outer state and hole tunneling to the QD are the slowest processes. This can be summarized as
\begin{equation}
  \frac{1}{\tilde T_1}\gg\tilde\gamma_a,\gamma_{c,d}\gg\gamma_{c,o},\gamma_{o,d}\:.
\end{equation}
Accordingly one can separate three sources of Kerr rotation noise: fluctuations of spin (SN), of QD occupation \mbox{(ON-d)}, and of outer-state occupation \mbox{(ON-o)}. We recall, that Kerr rotation is only present if the QD is in the charged state. The periods of the empty QD state can be either short, on the order of $\tilde\gamma_a^{-1},\gamma_{c,d}^{-1}$, or long, on the order of $\gamma_{c,o}^{-1},\gamma_{o,d}^{-1}$, when the hole is captured by the outer state and blocks hole capture to the QD. Naturally this leads to three different timescales in the Kerr rotation noise frequency spectra.

In order to find the steady-state occupation of the QD one should consider the kinetic equation for the occupancy of the outer state:
\begin{equation}
  \label{eq:no}
  \dot{n}_o=\gamma_{c,d}(1-n-n_o)-\gamma_{o,d} n_o\:.
\end{equation}
Bearing in mind Eqs.~\eqref{eq:kinetic} one finds the steady-state QD occupation:
\begin{equation}
\label{eq:nav}
  \bar n=\frac{\gamma_{o,d}\left(\gamma_{c,o}+\gamma_{c,d}\right)}{\gamma_{o,d}\left(\gamma_{c,o}+\gamma_{c,d}\right)+\left(\gamma_{o,d}+\gamma_{c,o}\right)\left(\lambda\lambda'\tilde T_1+\tilde\gamma_a\right)}. 
\end{equation}
The average occupancy of the outer state is simply given by $\bar n_o=\left(1-\bar n\right)\gamma_{c,o}/\left(\gamma_{c,o}+\gamma_{o,d}\right)$. These expressions along with Eq.~\eqref{eq:P} completely define the true (in contrast to quasi-) steady state of the QD. 

The correlation functions of fluctuations can be found from the quantum regression theorem~\cite{Landau.2000,Carmichael.1993}, which states that for $\tau\ge0$:
\begin{subequations}
  \label{eq:flucts}
\begin{equation}
  \frac{\d}{\d\tau}\braket{\delta\theta_K(0)\delta S_z(\tau)}=-\frac{\braket{\delta\theta_K(0)\delta S_z(\tau)}}{\tilde T_1}+\lambda\braket{\delta\theta_K(0)\delta n(\tau)},\\
\end{equation}
\begin{align}
  \frac{\d}{\d\tau}\braket{\delta\theta_K(0)\delta n(\tau)}=-\lambda'\braket{\delta\theta_K(0)\delta S_z(\tau)}- \nonumber \\
  \left(\tilde\gamma_a+\gamma_{c,d}\right)\braket{\delta\theta_K(0)\delta n(\tau)}-\gamma_{c,d}\braket{\delta\theta_K(0)\delta n_o(\tau)},
\end{align}
\begin{align}
  \frac{\d}{\d\tau}\braket{\delta\theta_K(0)\delta n_o(\tau)}= &-\gamma_{c,o}\braket{\delta\theta_K(0)\delta n(\tau)} - \nonumber \\
   &\left(\gamma_{c,o}+\gamma_{o,d}\right)\braket{\delta\theta_K(0)\delta n_o(\tau)}.
\end{align}
\end{subequations}
The fluctuations of Kerr rotation as follows from Eq.~\eqref{eq:Kerr} are simply given by $\delta\theta_K=C_s\delta S_z+C_n\delta n$, where
\begin{eqnarray}
  C_s=\frac{4\mathcal E\Delta\left(\gamma_1^2-\Omega_z^2/4+\Delta^2\right)}{\left(\gamma_1^2+\Delta_+^2\right)\left(\gamma_1^2+\Delta_-^2\right)}, \nonumber
  \\
  C_n=\frac{\mathcal E\Omega_z\left(\gamma_1^2+\Omega_z^2/4-\Delta^2\right)}{\left(\gamma_1^2+\Delta_+^2\right)\left(\gamma_1^2+\Delta_-^2\right)}.
\end{eqnarray}
Therefore the initial conditions for the system~\eqref{eq:flucts} can be found from Eqs.~\eqref{eq:avers} and
\begin{eqnarray}
  \braket{\delta S_z\delta n_o}& = & \braket{\delta n_o\delta S_z}  =  -\bar S_z\bar n_o,\nonumber
  \\
  \braket{\delta n\delta n_o} & = & \braket{\delta n_o\delta n}=-\bar n\bar n_o.
\end{eqnarray}

Taking into account the separation of timescales it is straightforward to solve Eqs.~\eqref{eq:flucts}:
\begin{subequations}
  \label{eq:corrs}
  \begin{eqnarray}
    \braket{\delta\theta_K(0)\delta S_z(\tau)}= & \left(\braket{\delta\theta_K\delta S_z}-\lambda\tilde T_1\braket{\delta\theta_K\delta n} \right)\e^{-\left|\tau\right|/\tilde T_1}\nonumber \\
    & +\lambda\tilde T_1\braket{\delta\theta_K(0)\delta n(\tau)},
  \end{eqnarray}
  \begin{eqnarray}
    \braket{\delta\theta_K(0)\delta n(\tau)} =& \left(\braket{\delta\theta_K\delta n}+\frac{\gamma_{c,d}}{\gamma_a'}\braket{\delta\theta_K\delta n_o} \right)\e^{-\gamma_a'\left|\tau\right|} \nonumber \\
    & -\frac{\gamma_{c,d}}{\gamma_a'}\braket{\delta\theta_K(0)\delta n_o(\tau)},
  \end{eqnarray}
  \begin{equation}
    \braket{\delta\theta_K(0)\delta n_o(\tau)}=\braket{\delta\theta_K\delta n_o}\e^{-\gamma_{c,o}'\left|\tau\right|},
  \end{equation}
\end{subequations}
with the three relaxation rates: 
\begin{eqnarray}
\label{eq:3rates}
1/\tilde T_1, \nonumber \\
\gamma_a'=\tilde\gamma_a+\lambda\lambda'\tilde T_1+\gamma_{c,d}, \nonumber \\
\gamma_{c,o}'=\gamma_{c,o}+\gamma_{o,d}-\gamma_{c,o}\gamma_{c,d}/\gamma_a'. 
\end{eqnarray}
In Eqs.~\eqref{eq:corrs} we have taken into account that the correlation functions are even functions of $\tau$. One can see, that at timescales $\gg \tilde T_1$ the spin correlator adiabatically follows the correlator of $\delta n$, which in turn at timescales $\gg \gamma_a'^{-1}$ similarly follows the correlator of the outer-state occupancy.

The Kerr rotation noise frequency spectrum is given by the Fourier transform of $\braket{\delta\theta_K(0)\delta\theta_K(\tau)}$:
\begin{equation}
  \label{eq:total_spectrum}
  \left(\delta\theta_K^2\right)_\omega
  =A_s\frac{\tilde T_1}{1+(\tilde T_1\omega)^2}+A_{d}\frac{\gamma_a'}{\gamma_a'^2+\omega^2}+A_o\frac{\gamma_{c,o}'}{\gamma_{c,o}'^2+\omega^2},
\end{equation}
where the coefficients determine the partial contributions to the Kerr rotation noise power spectrum
\begin{eqnarray}
  \label{eq:power_spectra}
  A_s & = & 2C_s\left(\braket{\delta\theta_K\delta S_z}-\lambda\tilde T_1\braket{\delta\theta_K\delta n} \right), \nonumber \\
  A_{d} & = & 2\left(C_s\lambda\tilde T_1+C_n\right)\left(\braket{\delta\theta_K\delta n}+\frac{\gamma_{c,d}}{\gamma_a'}\braket{\delta\theta_K\delta n_o} \right), \nonumber \\
  A_o & = & -2\left(C_s\lambda\tilde T_1+C_n\right)\frac{\gamma_{c,d}}{\gamma_{a}'}\braket{\delta\theta_K\delta n_o}.
\end{eqnarray}
One can see, that in general the spectrum indeed consists of three Lorentzian peaks centered at zero frequency with the HWHM given by $1/\tilde T_1$, $\gamma_a'$, and $\gamma_{c,o}'$, Eq.~\eqref{eq:3rates}, respectively.

In order to compare the analytical calculation with the experimental results one should take into account slow charge fluctuations in the QD environment, beyond the minimal model sketched in Fig.~\ref{fig:model}(a). These fluctuations lead to stochastic Stark shifts of the trion resonance frequency and can be accounted for by averaging the final results with a Gaussian distribution of the QD resonance frequency, $\exp\left[-\left(\omega_0-\bar\omega_0\right)^2/\Delta\omega_0^2\right]/\left(\sqrt{\pi}\Delta\omega_0\right)$, where $\bar\omega_0$ is the average trion resonance frequency and $\Delta\omega_0$ is the amplitude of its slow variations. 

\section{\label{s:dis}Discussion}

The theoretical calculation based on the presented model of spin and charge dynamics involves a number of physical quantities which can be determined by adjusting the free parameters of the theoretical model to the experimental results. The set of physical quantities is summarized in Tab.~\ref{tab:params} together with their respective values which yield a good agreement between the experimental and theoretical Kerr rotation noise spectra, as shown in Figs.~\ref{fig:spectra} and~\ref{fig:power}. Despite the large number of parameters to adjust the theoretical model, most of the physical quantities can be determined distinctly by examining the experimental data, as we discuss below.

First, the quantities related to the optical transition, namely the ``dressed'' homogeneous linewidth $\gamma_1$, the inhomogeneous broadening $\hbar\Delta\omega_0$ and the effective Zeeman splitting $\hbar\Omega_z$, can be determined from the noise power spectra, Fig.~\ref{fig:power}(b). The homogeneous linewidth determines the energetic splitting of the two maxima in the SN power spectrum, which is reproduced by the theory for a value of $\gamma_1 = 4.3\ \mu$eV. In the case of pure homogeneous broadening of the optical transition the SN power spectrum drops to zero at the trion resonance ($\Delta=0$).\cite{Wiegand.2018,Zapasskii.2013} Inhomogeneous broadening of the transition results in smearing of the dip in the average SN power spectrum whereat the finite minimum at zero detuning is a measure of the ratio between inhomogeneous and homogeneous broadening of the optical transition.\cite{Zapasskii.2013} Thereby we extract an inhomogeneous broadening of $\hbar\Delta\omega_0= 4.5\ \mu$eV which suggests that residual charge fluctuations in the solid-state environment of the QD are weak. We observe a significantly larger inhomogeneous broadening on the same sample where the QD density is higher.\cite{Dahbashi.2014} The main cause might be the higher degree of ionized acceptors since a larger number of holes are captured by the larger number of QDs.\footnote{A quantitative correlation between inhomogeneous broadening and QD density can not be given since only QDs with an optical transition resonant to the cavity mode are visible in our PL signal.}  

The effective Zeeman splitting between hole and trion spin states follows from the ratio of SN to ON power since the occupation noise strongly depends on the longitudinal magnetic field. The measurements performed at $B_z=31$~mT yield a Zeeman splitting on the order of the homogeneous linewidth, $\hbar\Omega_z=5.2\ \mu$eV.

Taking into account the quantities of the optical transition, the electron spin relaxation time $T_1^e$ directly follows from the widths of the measured SN spectra (blue dashed lines in Fig.~\ref{fig:spectra}(b)) with a value of 24~ns.
This is a typical longitudinal electron spin relaxation time for small $B_z$ and In(Ga)As QDs, which can be related to the hyperfine interaction~\cite{eh_noise,Smirnov.2017}. Note, that in the case of fast Rabi oscillations $\tilde T_1$ is a factor of two longer than $T_1^e$ (neglecting hole spin relaxation) since the trion state is occupied only half of the time on average. The Zeeman splitting and inhomogeneous broadening on the order of the homogeneous linewidth additionally increase the pseudospin relaxation time.

Interestingly, the formation of detuning-dependent dressed states leads to a spin polarization of the QD. The pseudospin orientation depends on the sign of detuning $\Delta$ and is given by Eq.~\eqref{eq:P}. The degree of polarization is shown in Fig.~\ref{fig:model}(b). One can see, that the polarization is significant and can reach almost $P=\pm50\%$ at $|\Delta|=\sqrt{\gamma_1^2+\Omega^2/4}\approx 5~\mu$eV.

Finally, we consider the quantities related to the charge dynamics. The sum (ratio) of Auger recombination rate $\gamma_a$ and direct QD-reoccupation rate $\gamma_{c,d}$ determines the width (amplitude) of the corresponding contribution in the Kerr rotation noise spectra (\mbox{ON-d} in Fig.~\ref{fig:spectra}(b)). The magnitude of the Auger rate in In(Ga)As QDs is well-known from recent measurements\cite{Kurzmann.2016,Wiegand.2018} to be on the order of $2\;\mu$s$^{-1}$. Together with a reoccupation rate of $\gamma_{c,d}= 0.7\;\mu$s$^{-1}$ the measured ON contribution related to direct QD recharging from the continuum is well reproduced by the theory in agreement with the result in Ref.~\onlinecite{Wiegand.2018}.
The narrow ON contribution related to the recharging channel via the outer state (\mbox{ON-o} in Fig.~\ref{fig:spectra}(b)) is similarly determined by the rates $\gamma_{c,o}$ and $\gamma_{o,d}$ which we find to be an order of magnitude lower than direct QD recharging with $\gamma_{c,o}=0.06\;\mu$s$^{-1}$ and $\gamma_{o,d}=0.03\;\mu$s$^{-1}$. The ratios $\gamma_a'/\gamma_{c,d}$ and $\gamma_{c,d}/\gamma_{o,d}$ define the occupancies of QD and outer state, which determine the noise power (area) of the respective contributions.

One can see, that despite the relative complexity of the model, its parameters can be reliably determined by comparison with the experimental results. However, the theoretical model can not account for a higher number of outer states and correlations of charges between them. As a result we obtain moderate agreement between theoretical model and experimental spectra at positive laser detunings (not shown) compared to the spectra at negative detunings where the agreement is very good (cf. Fig.~\ref{fig:spectra}(b)). In particular, the Stark shift of the QD resonance due to holes in different localized states can lead to the asymmetry of the noise spectra with respect to the laser detuning. On the other hand the asymmetry can also be caused by the interference of the light extracted from the cavity and reflected from the top Bragg mirror.\cite{Yugova.2009}
 
The most important parameters extracted from the analysis are the two reoccupation rates from the continu\-um into either the QD or into the nearby outer acceptor state, $\gamma_{c,d} = 0.7\;\mu$s$^{-1}$ and $\gamma_{c,o} = 0.06\:\mu$s$^{-1}$, respectively. They show that the capture cross section of the QD is about one order of magnitude larger than the capture cross section of the acceptor. Nevertheless, hole capture by the acceptor plays an important role since tunneling from the acceptor to the QD can be slow. In our specific QD, the hole tunneling rate $\gamma_{o,d}$ is $0.03\;\mu$s$^{-1}$, i.e., the acceptor is very close to the QD. However, $\gamma_{o,d}$ depends exponentially on the width of the tunneling barrier and can thus become extremely small in other QDs. Already in our case the average occupation of the QD by a hole reduces to about 30~\% (see Fig.~\ref{fig:model}(b)) for resonant driving of the trion state. In other QDs it is possible that neither spin nor occupation noise is observed since the occupation probability of the QD by a hole approaches zero for quasi-resonant driving of the trion resonance.

\section{\label{s:con}Conclusion}

We have shown that Kerr rotation noise spectroscopy is a powerful tool to study the hole occupation dynamics of a single QD which is in competition with a nearby ionized impurity. We have developed a theoretical model which explains at the same time spin, QD-occupation, and acceptor-occupation noise and yields a detailed quantitative understanding of the system. The fit of the presented model to the Kerr rotation noise frequency spectra at different detunings allows us to extract all relevant parameters of the spin and charge dynamics of the QD system. The parameters yield most importantly the two significantly different hole trapping rates from the continuum into either QD or acceptor which has a strong impact on the QD occupation.
The theoretical analysis reveals that the Zeeman splitting of the trion transition leads to strong spin pumping by linearly polarized light. Furthermore we have shown that the occupation probability of the QD by a hole is strongly reduced for resonant and quasi-resonant optical excitation of the trion since the reoccupation of the QD after an Auger process is rather slow in weakly p-doped samples. In view of spin-photon interfaces such a slow reoccupation is a drawback and gives reason for using more complex QD systems which ensure controllable hole reoccupation by tunneling contacts.

\begin{acknowledgments}

We thank K. Pierz (PTB) for providing the sample and M. M. Glazov (Ioffe Institute) for fruitful discussions. We acknowledge the financial support by the joint research project Q.com-H (BMBF 16KIS00107) and the German Science Foundation (DFG) (GRK 1991, OE 177/10-1). The theory was developed under partial support of the Basis Foundation and the Russian Science Foundation (Grant No. 14-12-01067).

\end{acknowledgments}

\renewcommand{\i}{\ifr}


%

\end{document}